\documentclass[]{spie}  

\usepackage{amsmath,amsfonts,amssymb}
\usepackage{graphicx}
\usepackage[colorlinks=true, allcolors=blue]{hyperref}

\title{Status of the PSF Reconstruction Work Package for MICADO@ELT}

\author[a]{Andrea Grazian}
\author[a]{Matteo Simioni}
\author[a]{Carmelo Arcidiacono}
\author[b]{Jani Achren}
\author[c]{Yann Clenet}
\author[d]{Yixian Cao}
\author[d]{Richard Davies}
\author[a]{Marco Gullieuszik}
\author[e]{Tapio Helin}
\author[f,g]{Daniel Jodlbauer}
\author[h,i]{Hanindyo Kuncarayakti}
\author[j]{Miska Le Louarn}
\author[h]{Seppo Mattila}
\author[k]{Fernando Pedichini}
\author[k]{Roberto Piazzesi}
\author[l,a]{Elisa Portaluri}
\author[e]{Aleksi Salo}
\author[m]{Gijs Verdoes Kleijn}
\author[a]{Benedetta Vulcani}
\author[f,g]{Roland Wagner}
\author[h,i]{Steven Williams}
\author[a]{Anita Zanella}
\author[n]{Werner W. Zeilinger}

\affil[a]{INAF Osservatorio Astronomico di Padova, Vicolo dell'Osservatorio 5,
I-35122, Padova, Italy}
\affil[b]{Incident Angle Oy, Capsiankatu 4 A 29, FI-20320 Turku, Finland}
\affil[c]{LESIA, Observatoire de Paris, Section de Meudon 5, place
Jules Janssen, F-92195 Meudon Cedex, France}
\affil[d]{MPE - Max-Planck-Institut für extraterrestrische Physik
Giessenbachstrasse 1, D-85748 Garching bei Muenchen, Germany}
\affil[e]{LUT University, P.O.Box 20, FI-53851, Lappeenranta, Finland}
\affil[f]{Industrial Mathematics Institute, Johannes Kepler University Linz,
Altenberger Strasse 69, 4040 Linz, Austria}
\affil[g]{RICAM - Johann Radon Institute for Computational and Applied
Mathematics, Altenberger Strasse 69, 4040 Linz, Austria}
\affil[h]{Tuorla Observatory, Department of Physics and Astronomy,
FI-20014, University of Turku, Finland}
\affil[i]{Finnish Centre for Astronomy with ESO (FINCA), Quantum,
Vesilinnantie 5, University of Turku, FI-20014 Turku, Finland}
\affil[j]{European Southern Observatory, Karl Schwarzschild Strasse 2,
D-85748 Garching bei Muenchen, Germany}
\affil[k]{INAF - Osservatorio Astronomico di Roma, Via Frascati 33, I-00078,
Monte Porzio Catone, Italy}
\affil[l]{INAF - Osservatorio Astronomico d'Abruzzo, Via Mentore Maggini,
I-64100 Teramo, Italy}
\affil[m]{University of Groningen, PO Box 72, 9700 AB Groningen,
the Netherlands}
\affil[n]{Department of Astrophysics, University of Vienna,
Tuerkenschanzstrasse 17, A-1180, Wien, Austria}

\authorinfo{Send correspondence to A.G. or M.S.
  \\ A.G.: E-mail: andrea.grazian@inaf.it, Telephone: +33 049 829 3465
  \\ M.S.: E-mail: matteo.simioni@inaf.it, Telephone: +39 049 829 3547}

\begin{document} 
\maketitle

\begin{abstract}
MICADO is a workhorse instrument for the ESO ELT, allowing first light
capability for diffraction limited imaging and long-slit spectroscopy
at near-infrared wavelengths. The PSF Reconstruction (PSF-R) Team of
MICADO is currently implementing, for the first time within all ESO
telescopes, a software service devoted to the blind reconstruction of
the PSF. This tool will work independently of the science data, using
adaptive optics telemetry data, both for Single Conjugate (SCAO) and
Multi-Conjugate Adaptive Optics (MCAO) allowed by the MORFEO
module. The PSF-R service will support the state-of-the-art
post-processing scientific analysis of the MICADO imaging and
spectroscopic data. We provide here an update of the status of the
PSF-R service tool of MICADO, after successfully fulfilling the Final
Design Review phase, and discuss recent results obtained on simulated
and real data gathered on instruments similar to MICADO.
\end{abstract}


\section{INTRODUCTION}
\label{sec:intro}  

MICADO (Multi-Adaptive Optics Imaging Camera for Deep Observations) is
the first light instrument of the European Southern Observatory (ESO)
Extremely Large Telescope (ELT) and it will be a workhorse facility
for Adaptive Optics (AO) diffraction limited imaging and long-slit
spectroscopy at near-infrared wavelengths \cite{micado2010}. We
provide here an update of the status of the point spread function
reconstruction (PSF-R) service tool of MICADO, after successfully
fulfilling the Final Design Review (FDR) phase, and discuss recent PSF-R
results obtained on simulated and real data gathered on instruments
similar to MICADO \cite{2020SPIE11448E..37S}.

A first-generation ELT instrument, MICADO will take high resolution
images of the Universe at near-infrared wavelengths. This makes MICADO
the ideal instrument for identifying exoplanets, but also for
resolving individual stars in local galaxies (up to Virgo and Fornax),
the star forming clumps in high-redshift galaxies, including the
dynamics of dense stellar systems, and investigating the mysterious
center of the Milky Way. The optimal scientific exploitation of MICADO
implies the detailed knowledge of the PSF of each acquired science
image or spectrum. The PSF-R service will support the state-of-the-art
post-processing scientific analysis of the MICADO imaging and
spectroscopic data, but it will not be available
for coronographic observations.
Indeed, the knowledge of the reconstructed PSF is
essential in order to evaluate the quality of the observed
astronomical data. Additional to quality evaluation, the reconstructed
PSF can be used for further improving the reduction of images and
spectra in the post processing phase.

The structure of this paper is the following: in Section
\ref{sec:method} we describe the methodology adopted for the PSF
reconstruction of AO-assisted diffraction limited instruments, e.g.
MICADO. The first results obtained by the MICADO PSF-R Team have been
highlighted in Section \ref{sec:results}. The size of telemetry data
required by the PSF-R Service every day is discussed in Section
\ref{sec:datarates}, while conclusions are drawn in Section
\ref{sec:conclusions}.

\section{METHOD}
\label{sec:method}

Different methods are available to obtain the PSF of AO-assisted
instruments: pure PSF Reconstruction (PSF-R), Hybrid, Calibrated,
Empirical, Adaptive methods. All these methods have been described in
detail in \cite{2020SPIE11448E..0AB}. MICADO will adopt as baseline
the pure PSF Reconstruction method, relying only on telemetry data,
without accessing the focal plane data (i.e. the scientific frames).

The motivations behind the choice of the Pure (or blind) PSF-R method
rely on different strategic points. For 35-40\% of the extragalactic
targets, a MICADO science frame will be void of point sources suitable
for standard PSF characterization (e.g. with empirical or
semi-empirical methods). Extragalactic deep fields are selected for
having as few stars as possible. In general, it is unlikely to have
interesting or well studied extragalactic targets in fields with many
stars per square arcmin. Extragalactic fields usually have no bright
stars suitable to build reliable empirical PSF model, nor reliable
empirical PSF at the position of each galaxy or rare objects
(e.g. high-z/lensed targets).

In case of crowded fields, it is difficult to recover the shape of the
PSF, especially on the wings, due to contamination from neighboring
stars. For these reasons, the Pure PSF-R approach is required when
extended targets cover most of the FoV.

The point spread function reconstruction (PSF-R) Application is a
deliverable of the MICADO project. The PSF-R Team of MICADO is
currently implementing, for the first time within all ESO telescopes,
a software service devoted to the pure (blind) reconstruction of the
PSF. This tool will work independently of the science data in the
focal plane, only using adaptive optics telemetry data of the MICADO
camera and spectrograph, both for Single Conjugate (SCAO) and
Multi-Conjugate Adaptive Optics (MCAO) allowed by the MORFEO module,
in post-processing analysis.

The PSF-R application will provide the reconstructed PSFs through an
archive querying system, using the Wave Front Sensor telemetry data
that MICADO will generate and save synchronously to each associated
science frame. The PSF-R Application of MICADO is designed to work
without accessing to the science data themselves, differently from
other PSF-R techniques that instead use the science data themselves to
optimize the parameters used during the reconstruction phase
\cite[e.g.]{2021SPIE11448E..2TN}. This allows the pure PSF-R tool to
be the most generally applicable, especially in the extra-galactic
cases where empty fields, devoid of bright stars, are routinely
observed.

The PSF-R Team will build up the software tools and architecture
needed to develop and test the PSF-R Service for MICADO, when the
instrument will be mounted on the bench, at the Preliminary Acceptance
in Europe (PAE) and later on at the ELT site in Cerro Armazones for the Final
Acceptance.

The pure PSF-R method has been described in detail in
\cite{1997JOSAA..14.3057V,2018JATIS...4d9003W,2020SPIE11448E..37S}, and
\cite{2022arXiv220901563S}.

\section{RESULTS}
\label{sec:results}

\begin{figure}[ht]
\begin{center}
  \includegraphics[height=6.0cm]{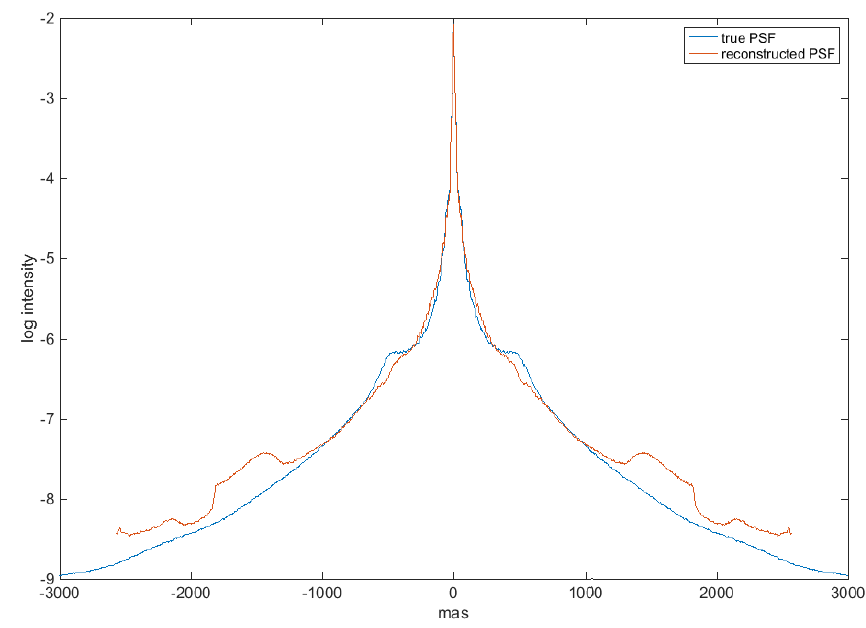}
  \includegraphics[height=6.0cm]{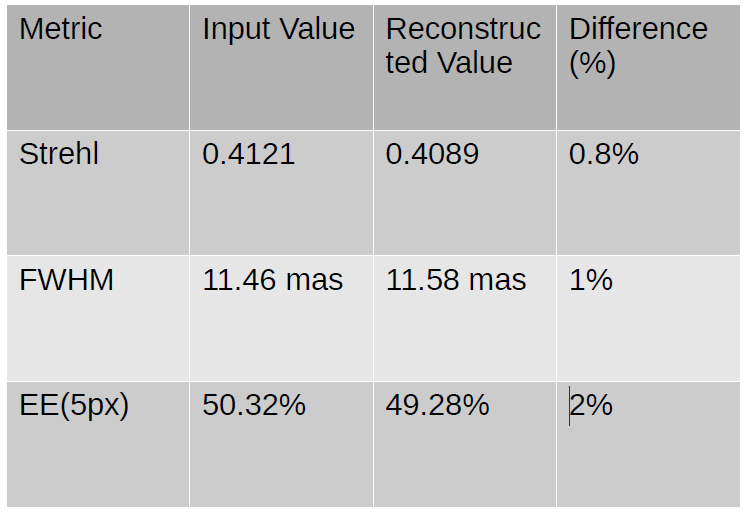}
\end{center}
\caption[datarate]
{\label{fig:scaopsfr}
The radial profile of the simulated (blue) and
reconstructed (orange) PSF. Accuracies of 1-2\% have been
reached in Strehl ratio, FWHM, and EE.}
\end{figure}

The pure PSF-R algorithms of MICADO have been tested using COMPASS
\cite{2016SPIE.9909E..71G} simulated data with a bright star (15th
magnitude), a Shack-Hartmann Wave Front Sensor (SH WFS) and the ELT M4
geometry. Strehl ratio (SR), FWHM, and encircled energy (EE) of the
simulated star are recovered by the pure PSF-R software at the 1-2\%
level (Fig. \ref{fig:scaopsfr}). Fig. \ref{fig:scaopsfr} shows the
radial profile of the simulated (blue) and reconstructed (orange) PSF.

Real data from the SOUL \cite{pinna_soul_2016} Adaptive Optic module
of the Large Binocular Telescope (LBT) have been used in order to
check that the SCAO Critical Algorithms of the MICADO PSF-R Service
can successfully process also real data from other instruments. The
first results show that the PSF-R algorithms are able to recover the
Strehl ratio (SR), FWHM and encircled energy profiles within a few
($\sim 1-7\%$) percent level \cite{2020SPIE11448E..37S}. Based on this
tests, we can confirm that the MICADO PSF-R Application has reached at
present a Technology Readiness Level (TRL) corresponding to TRL=7.
Further details on the PSF reconstruction of LBT/SOUL+LUCI data can be
found in this proceeding (Simioni et al. 2022).

The Final design Review (FDR) of the PSF-R software tool for MICADO has
been successfully accomplished in July 2021. This review has
been mainly focused on the development plan, the data flow,
the telemetry data rate, the end-to-end workflow, and the
development of the critical algorithms for PSF-R. No critical item has been
emerged during the FDR, indicating that the maturity level reached by the
MICADO PSF-R software is adequate for that milestone.

The scientific evaluation of the PSF-R tools is currently on-going for
extragalactic cases (e.g. galaxy morphology, clumps), as detailed in
this proceeding (Simioni et al. 2022). We have also developed tools
and a pipeline to evaluate how uncertainties on the reconstructed PSF
translate into uncertainties on scientific measurements (and
viceversa). In the future, the scientific evaluation of the PSF-R
products will be expanded to other science cases.

\section{TELEMETRY DATA RATES}
\label{sec:datarates}

The pure PSF-R software of MICADO relies only on WFS telemetry data,
saved at the highest temporal frequency, in order to better reproduce
the sudden variation of the atmospheric conditions.

\begin{figure}[ht]
\begin{center}
\includegraphics[height=10.0cm]{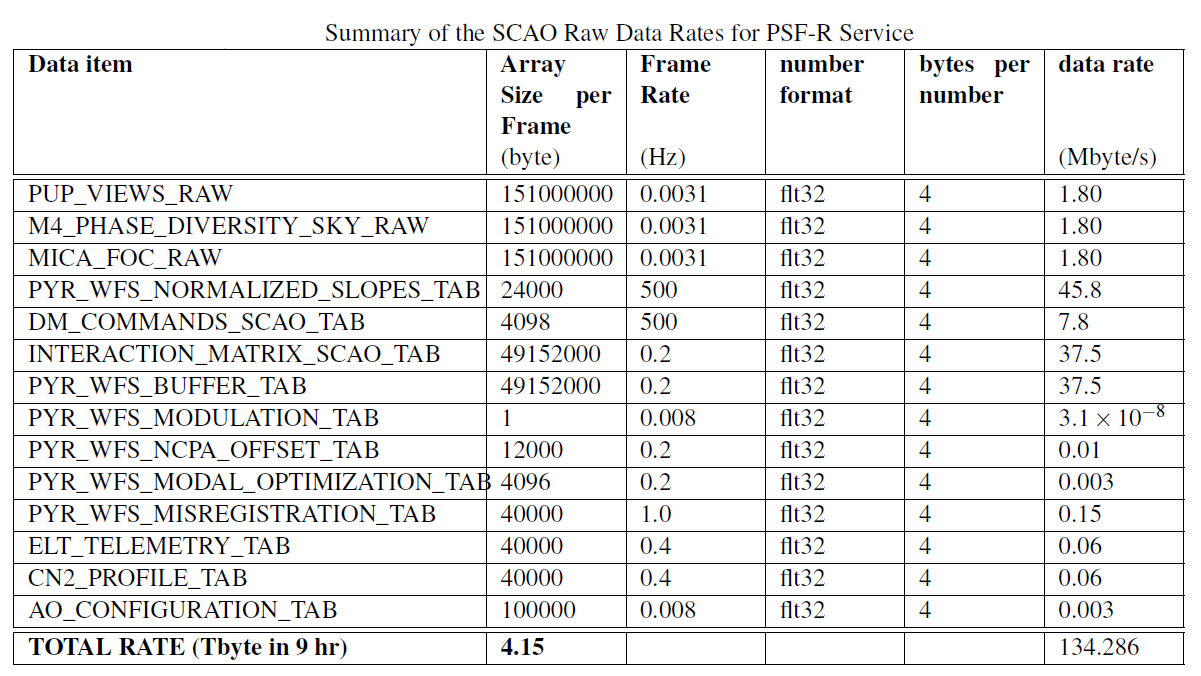}
\end{center}
\caption[datarate]
{\label{fig:datarate}
The expected data rates for the SCAO option of the PSF-R Service of MICADO.}
\end{figure}

The telemetry data rate of MICADO WFS is 4.2 Terabyte per night,
assuming 9 hours of observation, 100\% of open shutter time, control,
and interactive matrices saved at a frequency of 0.2 Hz. The telemetry
data rate is instead 2.1Tb per night, assuming 70\% open shutter time
plus control and interactive matrices saved at 0.1 Hz. The expected
data rate for telemetry is comparable or even less than the scientific
data rate. This amount of data rate is not a problem considering the
planned ESO/ELT infrastructure. Fig. \ref{fig:datarate} summarizes
the expected data rates in the SCAO case for the PSF-R Service of
MICADO. Preliminary calculations for the MCAO mode of MICADO indicate
that the expected data rate for MCAO is slightly lower than in SCAO
mode.

\section{CONCLUSIONS}
\label{sec:conclusions}

The present status of the PSF-R software tools of MICADO has been
summarised in this paper. In particular, a number of achievements have
been reached by the MICADO PSF-R team:
\begin{itemize}
\item
Accuracies of 1-7\% have been reached both with
simulated PSFs and with real LBT SOUL+LUCI data, as described in
\cite{2022arXiv220901563S}.
\item
A TRL=7 has been reached by the MICADO PSF-R
Service tools.
\item
There’s no issue for the ESO Raw Science Archive to
transfer and store all the WFS telemetry data produced by MICADO.
\item
No critical issue has been emerged during the FDR review, indicating that
the maturity level of the PSF-R tools is adequate for that milestone.
\item
No showstopper for PSF-R has been currently
identified, and the PSF-R Service is ready to meet the
First Light of MICADO at ELT in 2027.
\end{itemize}

The PSF-R software is a data intensive tool, since it requires
powerful computers and lot of disk space for storage. In the future,
the development of smart and efficient algorithms is required in order
to reduce the loading factor on the science users of ELT.

Further details on the PSF-R service for MICADO can be found in
this proceeding (Arcidiacono et al. 2022; Simioni et al. 2022).


\acknowledgments 
 
This work has been partly supported by INAF through the Math,
ASTronomy and Research (MAST\&R), a working group for mathematical
methods for high-resolution imaging.
This proceeding is in memory of our PSF-R team member Aleksi
Salo, who sadly passed away on June 11th 2022. Aleksi was
a friend and a talented young doctoral student with many
bright discoveries ahead. He will be greatly missed by his
family, friends, and colleagues. Farewell Aleksi, may the stars
shine upon you.

\bibliography{report}{} 

\begin{thebibliography}{1}

\bibitem{micado2010}
{Davies}, R., {Ageorges}, N., {Barl}, L., {Bedin}, L.~R., {Bender}, R.,
  {Bernardi}, P., {Chapron}, F., {Clenet}, Y., {Deep}, A., {Deul}, E., {Drost},
  M., {Eisenhauer}, F., {Falomo}, R., {Fiorentino}, G., {F{\"o}rster
  Schreiber}, N.~M., {Gendron}, E., {Genzel}, R., {Gratadour}, D., {Greggio},
  L., {Grupp}, F., {Held}, E., {Herbst}, T., {Hess}, H.~J., {Hubert}, Z.,
  {Jahnke}, K., {Kuijken}, K., {Lutz}, D., {Magrin}, D., {Muschielok}, B.,
  {Navarro}, R., {Noyola}, E., {Paumard}, T., {Piotto}, G., {Ragazzoni}, R.,
  {Renzini}, A., {Rousset}, G., {Rix}, H.~W., {Saglia}, R., {Tacconi}, L.,
  {Thiel}, M., {Tolstoy}, E., {Trippe}, S., {Tromp}, N., {Valentijn}, E.~A.,
  {Verdoes Kleijn}, G., and {Wegner}, M., ``{MICADO: the E-ELT adaptive optics
  imaging camera},'' in [{\em Society of Photo-Optical Instrumentation
  Engineers (SPIE) Conference Series}{\nolinebreak\hspace{0.1em}]},  {\em
  Society of Photo-Optical Instrumentation Engineers (SPIE) Conference Series}
  {\bf 7735},  77352A (Jul 2010).

\bibitem{2020SPIE11448E..37S}
{Simioni}, M., {Arcidiacono}, C., {Grazian}, A., {Clenet}, Y., {Davies}, R.,
  {Gullieuszik}, M., {Verdoes Kleijn}, G., {Pedichini}, F., {Wagner}, R.,
  {Ramlau}, R., {Zeilinger}, W.~W., {Vidal}, F., {Vulcani}, B., {Ragazzoni},
  R., {Sevin}, A., {Salasnich}, B., {Baruffolo}, A., {Busoni}, L., {Esposito},
  S., {Gendron}, {\'E}., {Piazzesi}, R., {Portaluri}, E., {Zanella}, A.,
  {Helin}, T., {Kuncarayakti}, H., {Mattila}, S., {Falomo}, R., and {Pinna},
  E., ``{MICADO PSF-reconstruction work package description},'' in [{\em
  Society of Photo-Optical Instrumentation Engineers (SPIE) Conference
  Series}{\nolinebreak\hspace{0.1em}]},  {\em Society of Photo-Optical
  Instrumentation Engineers (SPIE) Conference Series} {\bf 11448},  1144837
  (Dec. 2020).

\bibitem{2020SPIE11448E..0AB}
{Beltramo-Martin}, O., {Ragland}, S., {F{\'e}tick}, R., {Correia}, C., {Dupuy},
  T., {Fiorentino}, G., {Fusco}, T., {Jolissaint}, L., {Kamann}, S., {Marasco},
  A., {Massari}, D., {Neichel}, B., {Schreiber}, L., and {Wizinowich}, P.,
  ``{Review of PSF reconstruction methods and application to
  post-processing},'' in [{\em Society of Photo-Optical Instrumentation
  Engineers (SPIE) Conference Series}{\nolinebreak\hspace{0.1em}]},  {\em
  Society of Photo-Optical Instrumentation Engineers (SPIE) Conference Series}
  {\bf 11448},  114480A (Dec. 2020).

\bibitem{2021SPIE11448E..2TN}
{Neichel}, B., {Beltramo-Martin}, O., {Plantet}, C., {Rossi}, F., {Agapito},
  G., {Fusco}, T., {Carolo}, E., {Carl{\~A}}, G., {Cirasuolo}, M., and {Van Der
  Burg}, R., ``{TIPTOP: a new tool to efficiently predict your favorite AO
  PSF},'' in [{\em Adaptive Optics Systems VII}{\nolinebreak\hspace{0.1em}]},
  {\em Society of Photo-Optical Instrumentation Engineers (SPIE) Conference
  Series} {\bf 11448},  114482T (Jan. 2021).

\bibitem{1997JOSAA..14.3057V}
{Veran}, J.~P., {Rigaut}, F., {Maitre}, H., and {Rouan}, D., ``{Estimation of
  the adaptive optics long-exposure point-spread function using control loop
  data.},'' {\em Journal of the Optical Society of America A}~{\bf 14},
  3057--3069 (Nov. 1997).

\bibitem{2018JATIS...4d9003W}
{Wagner}, R., {Hofer}, C., and {Ramlau}, R., ``{Point spread function
  reconstruction for single-conjugate adaptive optics on extremely large
  telescopes},'' {\em Journal of Astronomical Telescopes, Instruments, and
  Systems}~{\bf 4},  049003 (Oct. 2018).

\bibitem{2022arXiv220901563S}
{Simioni}, M., {Arcidiacono}, C., {Wagner}, R., {Grazian}, A., {Gullieuszik},
  M., {Portaluri}, E., {Vulcani}, B., {Zanella}, A., {Agapito}, G., {Davies},
  R., {Helin}, T., {Pedichini}, F., {Piazzesi}, R., {Pinna}, E., {Ramlau}, R.,
  {Rossi}, F., and {Salo}, A., ``{Point spread function reconstruction for
  SOUL+LUCI LBT data},'' {\em JATIS accept.} ,  arXiv:2209.01563 (Sept. 2022).

\bibitem{2016SPIE.9909E..71G}
{Gratadour}, D., {Ferreira}, F., {Sevin}, A., {Doucet}, N., {Cl{\'e}net}, Y.,
  {Gendron}, E., {Lain{\'e}}, M., {Vidal}, F., {Brul{\'e}}, J., {Puech}, M.,
  {V{\'e}rinaud}, C., and {Carlotti}, A., ``{COMPASS: status update and long
  term development plan},'' in [{\em Adaptive Optics Systems
  V}{\nolinebreak\hspace{0.1em}]},  {Marchetti}, E., {Close}, L.~M., and
  {V{\'e}ran}, J.-P., eds., {\em Society of Photo-Optical Instrumentation
  Engineers (SPIE) Conference Series} {\bf 9909},  990971 (July 2016).

\bibitem{pinna_soul_2016}
Pinna, E., Esposito, S., Hinz, P., Agapito, G., Bonaglia, M., Puglisi, A.,
  Xompero, M., Riccardi, A., Briguglio, R., Arcidiacono, C., Carbonaro, L.,
  Fini, L., Montoya, M., and Durney, O., ``{SOUL}: the {Single} conjugated
  adaptive {Optics} {Upgrade} for {LBT},'' in [{\em Adaptive {Optics} {Systems}
  {V}}{\nolinebreak\hspace{0.1em}]},   {\bf 9909},  99093V, International
  Society for Optics and Photonics (July 2016).

\end{thebibliography}
\bibliographystyle{spiebib} 

\end{document}